\documentclass[12pt]{article}
\usepackage[T1]{fontenc}
\usepackage[cp1250]{inputenc}
\usepackage{eepic}
\usepackage{graphicx}

\begin{document}

\begin{center}

{\bf MODEL OF WEALTH AND GOODS DYNAMICS IN A CLOSED MARKET}\\

 Marcel AUSLOOS$^a,\footnote{Corresponding author. E-mail address:
marcel.ausloos@ulg.ac.be}$ and Andrzej P\c{E}KALSKI$^b$ \\

\end{center}
{\it $^a$ SUPRATECS, B5, Sart Tilman Campus, B-4000 Li$\grave
e$ge, Euroland,\\
$^b$ Institute of Theoretical Physics,
University of Wroc{\l}aw, pl. Maxa Borna 9, PL-50-204 Wroc{\l}aw,
Poland }

%\date{\today} \\

%\preprint
%\draft

%\maketitle

\begin{abstract}
A simple computer simulation model of a closed market on a fixed
network with free flow of goods and money is introduced. The model
contains only two variables :  the amount of goods and money
beside the size of the system. An initially flat distribution of
both variables is presupposed. We show that under completely
random rules, i.e. through the choice of interacting agent pairs
on the network and of the exchange rules that the market
stabilizes in time and shows diversification of money and goods. 
We also indicate  that the difference between poor and rich agents
increases for small markets, as well as for systems in which money
is steadily deduced from the market through taxation.
   \end{abstract}
\vskip 0.5cm


   \vskip 0.5cm

   \section{Introduction}
   Numerous works have inspired physicists and guided them towards the
   studies of financial markets. We could start with Bachelier (1900)
   \cite{bachelier}
   who introduced what shall perhaps remain the simplest and most
   successful model for price variations
   based on Gaussian
   statistics \cite{jura}.
 On the other hand,  Pareto (1897)
   \cite{pareto} had observed that large price
   variations  seem to be described by a curve interpolating
   between two power-laws.

 For an econophysicist using statistical mechanics  as primary support of
 investigations, nature seems to work as a  ``self-organized system''.
 \cite{Bak}
 ``Scaling laws'' are everywhere, but rarely contain cycles. Nevertheless it
is
understandable that physicists try to convince themselves that  there is some
punctuated equilibrium in economy, i.e. a market can be in
a metastable equilibrium.

  A fundamental question is that of `'value''; not only value of money,
    but more generally the value of goods
      \cite{Hallegatte}. In fact,
 in extreme situations, prices may not represent actual values. For instance,
in the aftermath of a natural disaster, the amount of destruction - measured
using pre-disaster prices - is a poor proxy for actual welfare losses,
because of reconstruction constraints and interactions with basic needs.
We will not debate on whether there is a
        real value of goods here  \footnote{No  need to recall the price of
    tulips in 1625-1635
      \cite{tulipomania}.}, but will consider whether
        in absence of such a notion a ''free'' market can exist, and if so what
would be
  its characteristics.

    Markets surely started  through barter,
       but later exchange of goods needed money for a surviving economy
In the real world, money seems needed for a  functioning of the economy.  Changes in
          the quantity of money may affect this functioning through inflation
          and deflation
          processes. However, there is strong
          empirical evidence that changes in the quantity of money affect
          the true value of goods, at least on the short term and modify the
          $risk$ faced by  agents,
          see e.g. \cite{r3}. Moreover we all complain about taxes,
          not only on income and capital, but also about so called VAT,
          imposed on any transaction. We will not debate their value,
          but it is of interest to examine whether taxes influence the
          characteristics of very simple exchange markets.

    Let us also, for the sake of perspective,
refer to a few previous reports on the question of wealth and values within physics
research work.  Models of prices are too numerous to be recalled here. Models of values
            and wealth are sometimes quite elaborate. Since we wish to devise a basic algorithm
            we only recall here below the following ones.

       Pianegonda et al.
         \cite{pianegonda1,pianegonda2} in a series of papers
       introduced one dimensional models of wealth redistributions.
It was of interest to
 compare things when the commercial links were frozen with the case in which the economical agents can
choose their commercial partners at each time step.
          Das and Yarlagadda
             \cite{cm0304685} mathematically analyzed a simple market model
         where trading at each point
         in time involves only two agents with the sum of their
money being conserved
and with neither parties coming out with negative money after the interaction
process. The exchange involved random re-distribution among the two players of
a fixed fraction of their total money.
   Chakraborti
      \cite{r0205221} studied the distributions of money  for
      different types of monetary transactions.
     He examined
    a   model  in which
    the agents invest equal amounts of money in each transaction.
Chatterjee \cite{Chatterjee}  studied the Pareto law appearance in
a kinetic  model of market  with  random  saving  propensity, A
pairwise exchange model using a slightly different exchange rule
has been published by Slanina  \cite{Slanina}. He also introduced
the possibility of money entering or leaving at each exchange
point.

Following the work by , Patriarca, Chakrabarti and Kaski (PCK)\cite{CPK}, 
Richmond
et al. managed to solve their collision model analytically and showed
that as formulated, it does not account for the observed data. The
Pareto exponent for the CPK exchange model is exactly one. Values
greater than 1 are not admitted by the model without introducing
additional complexity such as memory, breaking the conservation of
money during the exchange process or modifying the exchange rule as
indeed was done by Slanina. CPK and coworkers only did a numerical simulation and concluded 
that
because they got a 'Pareto exponent' just above one (they actually got
1.03) the model could account for the real data. The  papers \cite{Richm1,Richm2,Richm3} show
'au contraire', that it could not!! In fact, the Generalised Lotka Volterra solution is well able to
fit the real data.  As a result it is believed that the CPK exchange model  may
be abandoned - or at least require serious modification.  Some
economists also have strong philosophical objections to this
particular exchange model.

Recently Manolova et al.
\cite{Manolova1,Manolova2} have  studied an economy consisting of
a finite number of agents located on a one-dimensional torus. Each
agent buys a set  of goods from its nearest (left) neighbor,
produces some other goods, sells goods to its nearest (right)
neighbor and consumes the rest. Thus, there are two opposite
fluxes on the torus - goods and money.
 Each
  agent attempts to maximize a risk aversion utility function
  of the expected produced and consumed quantities w.r.t. its production and
  exchange decision. Information is local - each agent knows only the
history of its own buy, sell, and production decisions. The
individual agents plan over a revolving time horizon of two
periods. Thus, in each period, the agents must form anticipations
about what will happen in the next period (mean and variance),
based on their past experiences.
  The authors investigated the transitory and permanent impact of
  local or global
  injections of money on the dynamics of exchanged quantities, prices,
  and individual welfare,
  and the mechanisms that explain this evolution.

The distribution of wealth among the members of such a society is
often assumed to result from one fundamental mechanism:  trade.
More sophisticated models assume two mechanisms, like  trade and
savings \cite{r0205221} or trade and investment \cite{r0403045}.
Others take into account good productions like in ref.
\cite{sneppendonangelo}: it was shown that the dynamics of
exchange value in a system composed of many interacting agents
exhibit cooperative emergence and collapse of global value for
individual goods. Several extensions have been proposed and
published.

  It is still apparently unclear whether a more simple
constraint leads to wealth redistribution, or to huge disparity in the society.
It is of interest to use similar ideas, as those of previously quoted authors, considering  agents
  as interacting
  randomly, e.g. around a table, still imposing a constant amount of money and yet a constant
  amount of goods, the exchange involving random re-distribution between
  two agents of a  fraction of their total money and goods,
  with constraints to be defined in Sect.2.

  The goal of such a basic approach is  thus to devise the most basic model of exchanges
  wealth redistribution under the most simple rules, and value definition, i.e.
  when there is no anticipation, nor utility (like saving) constraint.
  In section 3, we will outline the Monte Carlo steps done for implementing the
model. We present a few results in Sect.4, - it will be seen that one can imagine a
society made of two groups, the merchants, having goods (and money), the bankers having money (and goods). We end with
conclusions in Sect.5.

\section{Model}
We present here a very simple model of goods and money flow. In
the model we consider a constant number, $N$, of agents,  each
possessing at the beginning  the same amount of goods, called here
apples, and money (e.g. {\it euro}). At each step an agent is
arbitrarily selected  to look randomly for a partner from whom she
may want to buy  apples. With a 50 \% probability the decision
could be either {\it yes, I want to buy}, or {\it no, I do not
want to buy}. The decision does not depend neither on previous
decisions nor on any other factor. It is completely random. The
details are explained in the algorithm given below. Then it is
determined randomly how many
 apples she would like to buy from such the partner, and
 how much money she is willing to pay. Then the partner is
 checked whether she has enough  apples and if so, whether
 she wants   to sell. When both answers are positive, the transaction
 is made and the apples and money are appropriately transferred.
In such a way there is no predetermined direction of neither flow.
The total amount of money and goods are conserved throughout the
process. The same is for the number of agents. Even if somebody
sold all  her apples and run out of money, she still stays in the
game, although rather as an observer, since she can neither buy
nor sell.

When we agreed that, like in the Manolova et al. paper
\cite{Manolova1}, agents are  pinpointed and the exchange is done
only between nearest neighbors, we  found that the dynamics
stopped quickly,
 since if close to a rich agent there is a very poor one,
 no trading is possible. Therefore in our model there is no fixed link
 among agents (or spatial position) and the underlying network of
 connections is a dynamical one.

The algorithm is the following:
\begin{enumerate}
\item  chose randomly  agent $j$,
\item check whether she has money and determine, with a 50 \% probability
(flipping a coin),  whether she wants to buy,
\item determine how many apples  (from 1 to 5, the maximum we allowed )
and for how much (random fraction of the total amount of money she
has),
\item find,  randomly, another agent  $k$,
\item check whether she has enough apples to sell to the agent
$j$,
\item if so, determine whether she wants to sell, again by tossing a
coin,
\item when positive, pass the apples from $k$ to $j$ and money from $j$  to
$k$,
\item when $N$  times the choice 1 has been made, one Monte Carlo Step (MCS)
has been completed.
\end{enumerate}

During the simulation we have recorded the following quantities:
\begin{itemize}
\item  amount of apples sold
\item amount of money transferred
\item number of transactions
\item maximum of apples possessed
\item maximum of money possessed
\item distribution of apples
\item distribution of money
\end{itemize}

In so doing we consider that the economy is bipolar : there are
''more merchant-like agents", with their distribution and status
characterized by the number of apples they own, and  more
banker-like agents characterized by the amount of money. Of
course, bankers have (buy or sell) apples and merchants own some
money as well, beside having apples.

\section{Results}

We have performed our simulations with the following values of the
parameters:  the total number of agents $N$ = 60, initial amount
of money per agents (the same for all) = 50 (``EUR''), initial
amount of apples, also 50. The simulations were performed till
10000 MCS and the results averaged over 500 independent runs,
which gave us quite good statistics. We have checked how robust
the results are with respect to the changes in the size of the
market, hence the number of agents, and the amount of goods
(apples) and money attributed at the beginning to each agent.

\begin{figure}
\begin{center}
\includegraphics[scale =0.6]{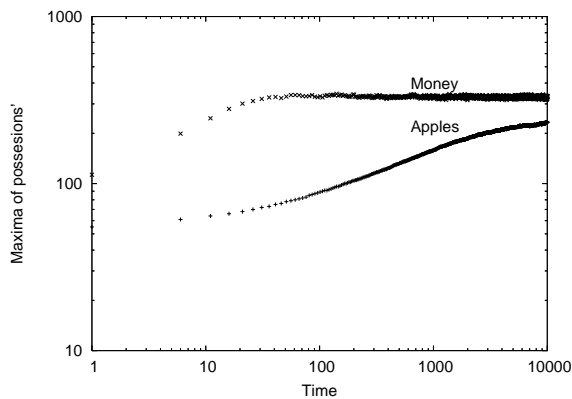}
\caption{Time dependence of the maximum amount of money and apples
 owned by one person. Logarithmic scales} \label{maxima}
\end{center}

\end{figure}

The time dependence of the maximum amount of apples and money  is
presented in figure \ref{maxima} on a doubly logarithmic scale. It
is seen that although   asymptotically   both curves are flat, the
richest agent arrives at her maximum faster than her colleague
having the maximum number of apples. This effect is most  probably
due to the fact that in order to slow down the dynamics we allowed
at most 5 apples to be sold in a single transaction. On the other
hand we did not put any limit on the price, which is restricted
only by the amount of money in the pocket of the buying agent.

Although the price is randomly determined by the seller and also
randomly accepted, or not, by the buyer, the fluctuations on the
number of  transactions per time, the number of exchanged apples
and amount of money, show only little scatter and remain at
approximately the same levels (11, 11, 35 respectively) throughout
the simulations.

\begin{figure}[here]
\begin{center}
\includegraphics[scale=0.6]{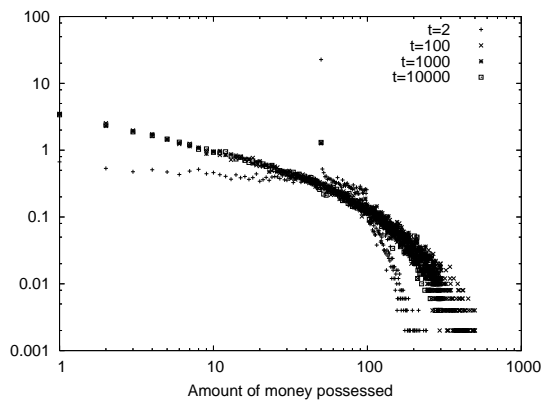}
\caption{Distribution of money in several time moments, shown on a
log-log scale} \label{di-m-log}
\end{center}

\end{figure}

\begin{figure}
\begin{center}
\includegraphics[scale=0.6]{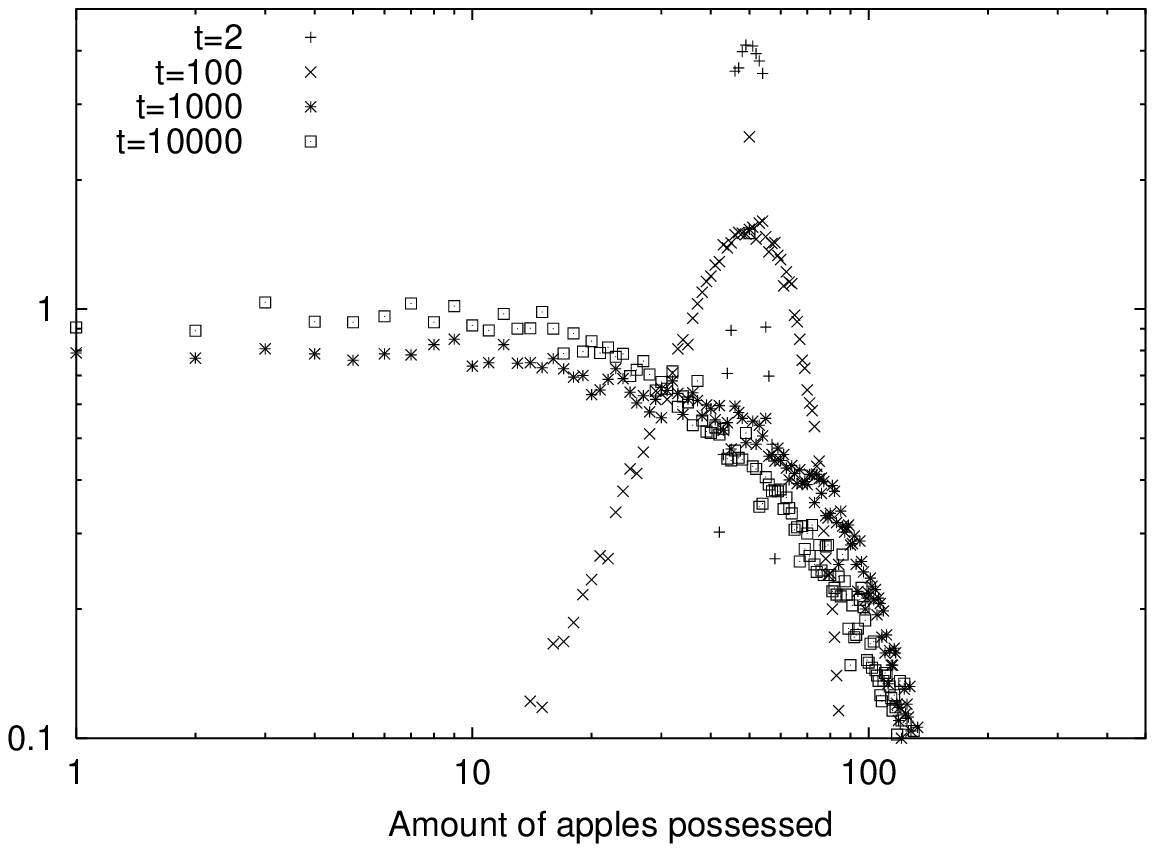}
\caption{Distribution of apples in several time moments, shown on
a log-log scale.} \label{di-a-log}
\end{center}

\end{figure}

The distribution of wealth, shown in figure \ref{di-m-log},  as
well as the distribution of goods (figure \ref{di-a-log}) show
that the initial distributions peaked at the values 50, spread
quite fast and after some 1000 MCS there is already some amount of
agents owing many apples and  others becoming quite rich. It
should be noted that the distributions are neither symmetric, nor
are they similar one to another. The distribution of wealth shows
many agents possessing below 20  ``EUR'' and just a few (the rich
bankers) having more than 140. The distribution of apples is not
so drastic, but again the majority of agents has below 20 apples;
there are few (rich) merchants.  This asymmetry with respect to
the wealth distribution is the results of our ``socialist''
restriction on the maximum of apples sold in one transaction.

This log-log plot  (Fig. \ref{di-m-log}) recovers the findings  of
Pareto for price distributions,
   \cite{pareto} i.e.,  the (money) wealth
   variations  seem to be described by a curve interpolating
   between two power-laws with some crossover. What can be pointed out is the
relative stability of the crossover point, and the asymptotic
nature of the power laws, respectively -0.5 and -2.0 at low and
high (money) wealth.  The (good) value  distribution  (Fig.
\ref{di-a-log}) i s different from the (money) wealth, at low
value, due to the type of market we impose. However a similar -2.0
asymptotic power law is found for the large amount of goods in
possession of the (rich in goods) agents.

\begin{figure}
\begin{center}
\includegraphics[scale=0.7]{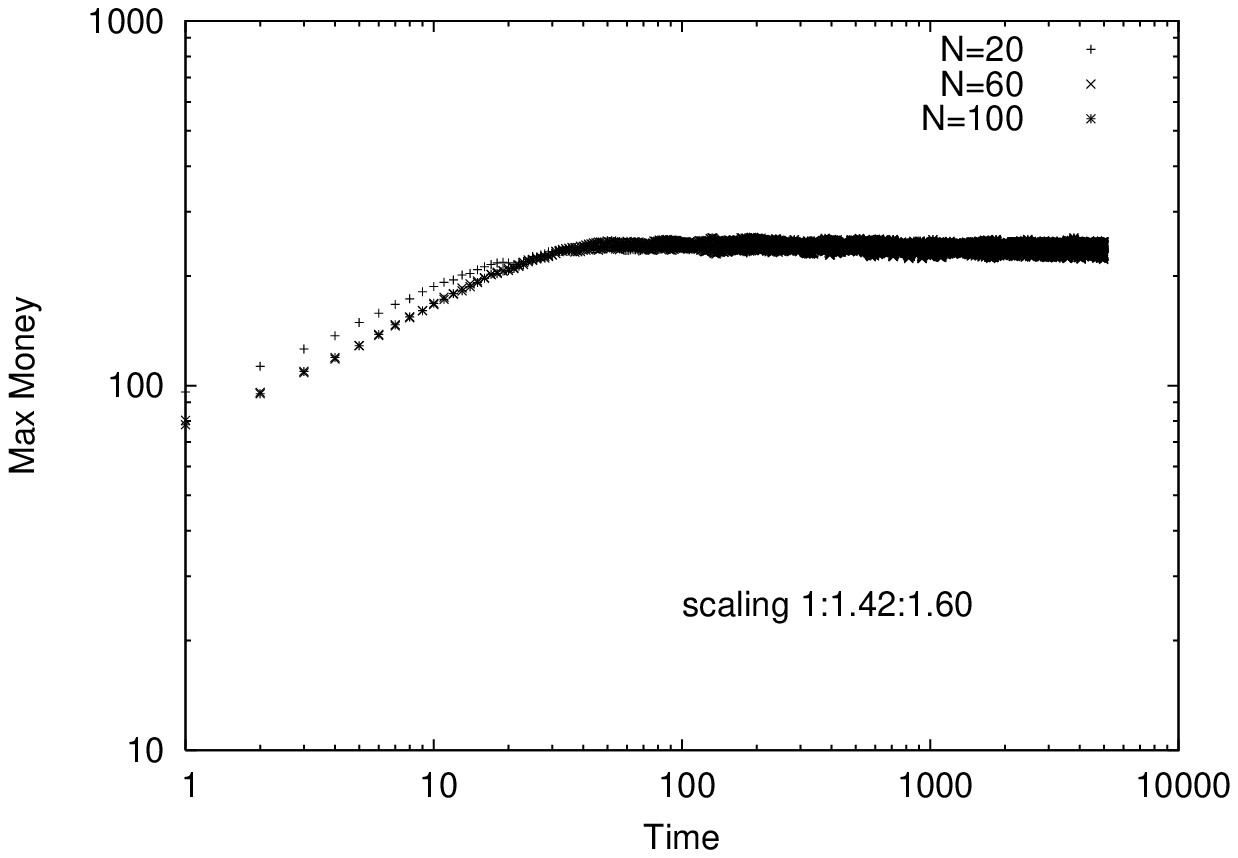}
\caption{Time dependence of the maximum number of apples in one
hand,  for so called small, medium and large market; log-log
scales.} \label{scal1}
\end{center}

\end{figure}

\begin{figure}
\begin{center}
\includegraphics[scale=0.7]{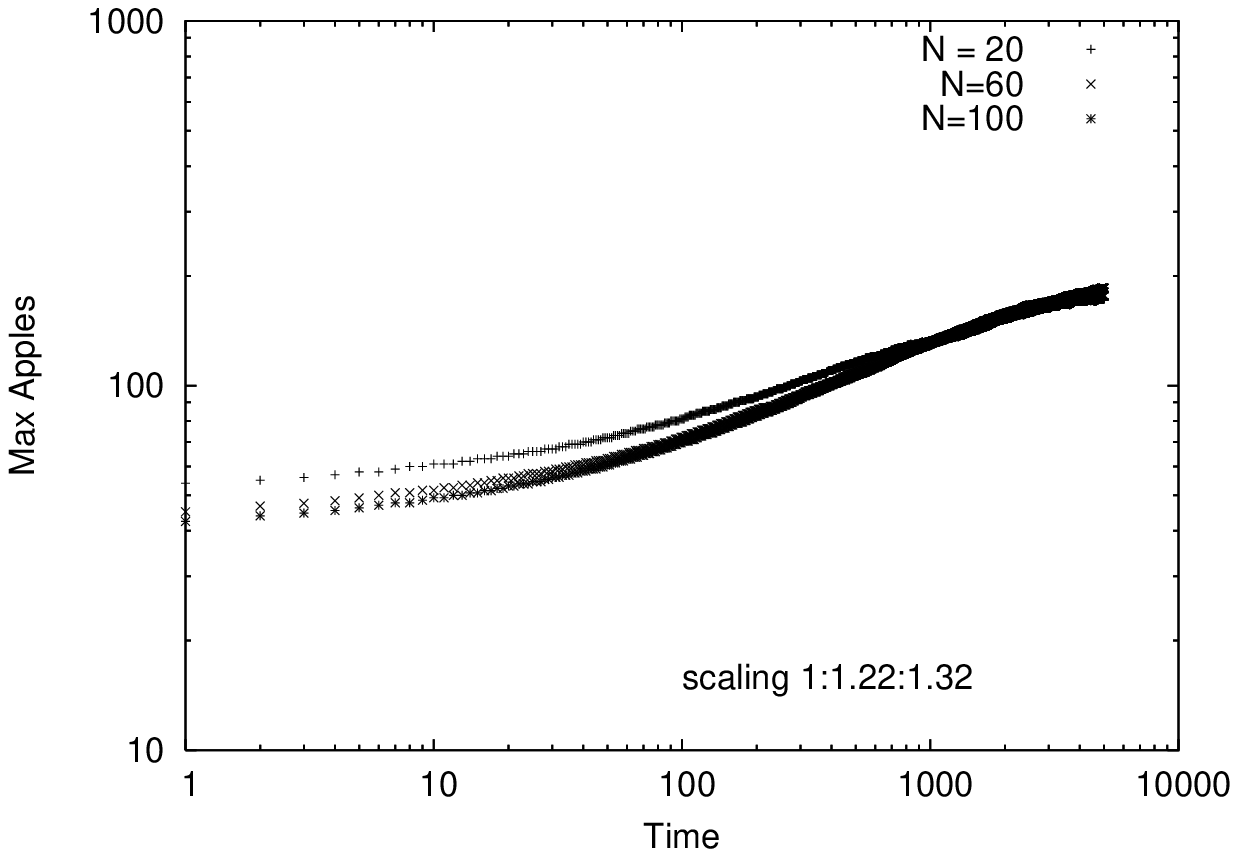}
\caption{Time dependence of the maximum amount of money in one
hand,  for so called small, medium and large markets; log-log
scales} \label{scal2}
\end{center}

\end{figure}

We have been also interested how the  results generated by our
simple model  change when we allow for a larger amount of agents.
Therefore we changed the number $N$  of agents to 20  and 100. Most
interesting would be the time development of the maximum of apples
and money, since for the average price and the distributions one
would expect the same behavior, regardless of the size of the
market. As could be seen from figures \ref{scal1} and \ref{scal2},
the curves, after appropriate scaling, are quite similar. In the
figure \ref{scal1} the data for $N$ = 60 were divided by 1.22 and
the data for $N$ by 1.32, while in the figure \ref{scal2} the
ratios were 1.42 and 1.60, respectively.

Finally, we decided to check if our model is sensitive  to some
modifications, therefore we introduced taxes. At each transaction
a certain amount of money paid goes not to the seller but to the
``state'', hence disappears from the market, which is no longer
closed, as far as money is concerned.
Notice that the ref.\cite{Richmond2} paper
 on the Lotka Volterra approach is also an exchange model
where the so-called autocatalytic term allows the redistribution or extraction of money from all agents simultaneously.
In this sense it can mimic a tax.

\begin{figure}
\begin{center}
\includegraphics[scale=0.7]{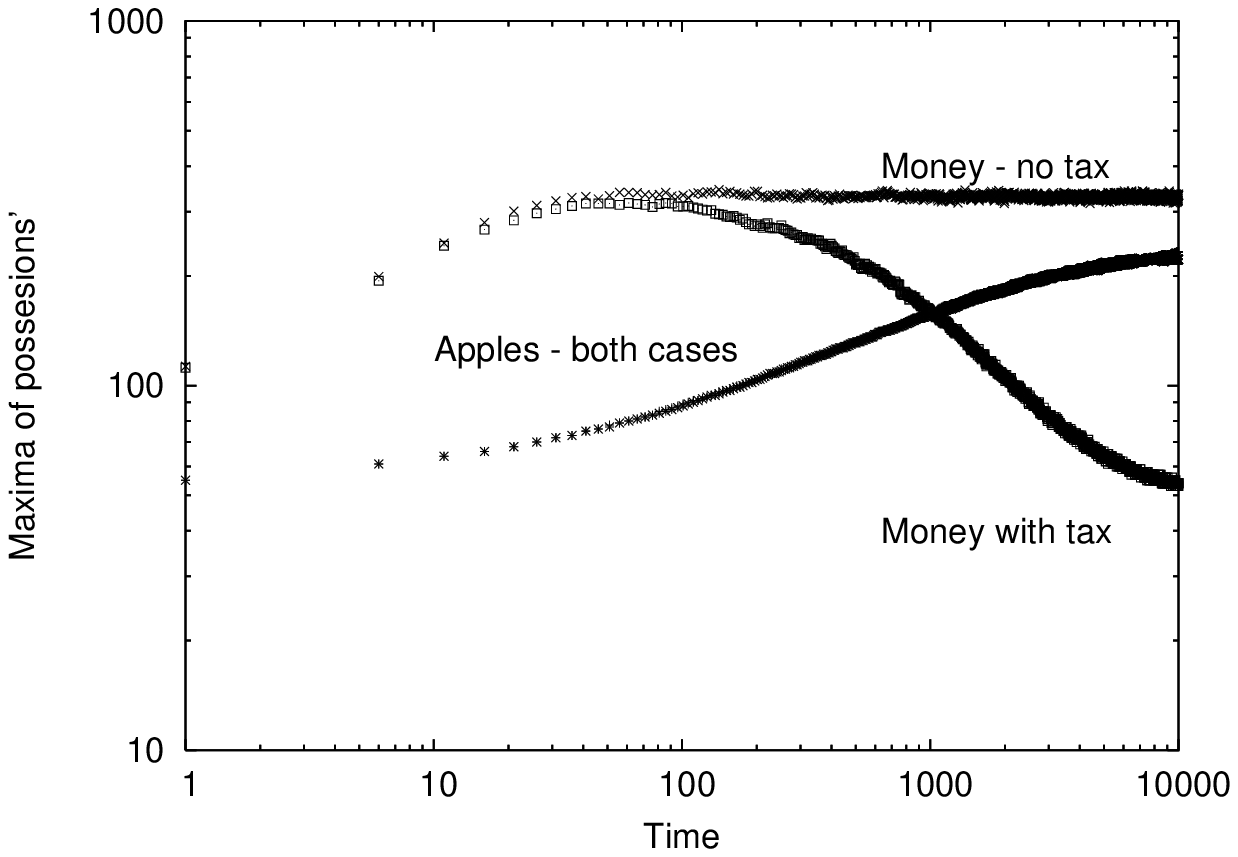}
\caption{Maximum of apples and money owned in the cases without
and with 1\% tax; log-log plot} \label{maxima3}
\end{center}

\end{figure}

\begin{figure}
\begin{center}
\includegraphics[scale=0.7]{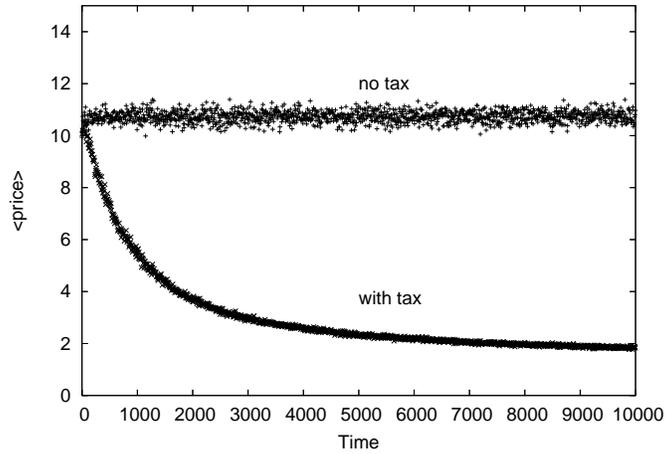}
\caption{Average price per apple in the cases without and with tax; normal scales}
\label{price3}
\end{center}

\end{figure}

 For the present
considerations we invented a flat tax rate of 1 \% on the money
which is exchanged. The amount of apples remains constant. The
changes in the results are presented in figures \ref{maxima3}  and
\ref{price3}. The remaining characteristics, like the number of
transactions, are the same as in the closed market. In the figures
we present data for the case of $N$ = 60 agents, each having at
the beginning 50 apples and 50 euros. As seen, the effect of taxes
is not immediately felt, - as often ! The maximum number of
apples owned by the best merchants increases with time, - it
accelerates when taxation is coming into play, on the richest
agents, for levelling off, when the taxation effect is slowing
down.

\begin{figure}
\begin{center}
\includegraphics[scale=0.7]{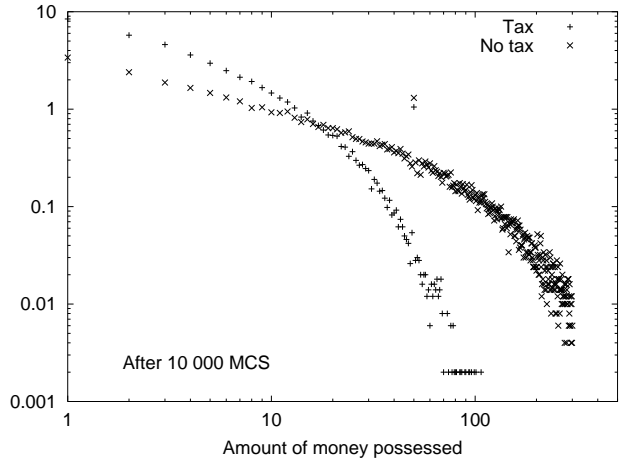}
\caption{Final distribution of money in the case with  and without
taxes on a log-log plot. $N$ = 60.} \label{distr-m-l-unl}
\end{center}
\end{figure}

Figure \ref{distr-m-l-unl} shows the effect of taxation  on
the distribution of wealth in the ``society''. It is clear that
taxation, which in our model reduces the amount of money
circulating in the system, has similar effect like reducing the
initial amount of money received by each agent. However this is
not seen at the initial stages of the simulation and the
distributions of wealth after, say 1000 MCS, show no essential
differences between the case {\it with taxes}  and {\it no taxes}.
After 10000 MCS there were no rich agents (having more than 100 euros) in the system with taxes and there were relatively many having just below 5 euros. We may therefore conclude that in the system with taxes the ''difference'' between the richest and the poorest agent is smaller
than in the system without taxes, - quite natural indeed. Also the average price per apple (figure \ref{price3}) goes down significantly. This however does not imply that in a system with taxes  the agents are, on average, better off. As shown in figure \ref{distr-m-l-unl}, there are more poor agents in that case. It should be also noticed that in our simple model the tax money is dissipated, it simply disappears from the system, without being reintroduced into it in any way.

 The
distributions of apples, even in the final stage, are quite
similar in the two cases.

     \section{Conclusions}
 We have presented a very simple model of a closed marked with free
exchange of goods and wealth.  The model contains just the
following parameters: size of the market (number of agents) and
the amount of goods and money owned initially by each agent.  The
role played
 in the dynamics by each of these factors has been investigated and discussed.
 The role of initial conditions has not been examined: e.g.   we have also
 assumed an initially
 uniform distribution. We have imposed completely random rules, i.e. the amount
 of goods and the accepted price, as well as the trade decisions have no relation
 to trends,
 previous activity, anticipation, savings,  etc.  Also partners for trading
 are chosen randomly among all agents. This is found to be very relevant in view of
maintaining a dynamic economy. This does not require money or good injections
as in Manolova et al. work. \cite{Manolova2},
as often thought of ''politically speaking'' as well. Of course, there
is a catalytic effect, in imposing inflation,
as qualitatively seen in comparing our results with theirs.

 Such a free market stabilizes itself pretty soon, -- the average price,
 number of transactions, number of goods sold and money paid for them,
 become asymptotically fixed in time.  In some sense  there is an
 intrinsically generated (''self-organized'') utility function. The
 distribution of money and goods shows a stratification of the ``society'':
  there are poor and wealthy agents; some have money, others own goods.
  This division is deeper in ``shallow''
 markets, i.e. when there is less money and goods circulating.
 Apart from that, many other characteristics of the market
 (number of transactions, average price, etc) have similar behavior and do
 not depend in a qualitative way on the size of the market. Most of the
 results appear ''quite natural''.

Considering characteristic exponents of power law tails for wealth distributions, 
 there seems
to be a trend away from Pareto's values of 1.5 to slightly higher
values. It can be wondered \cite{Richmond3} if this represents the
impact of socialism since a higher slope value is associated with
greater redistribution of money, as we found, whereas a low value of the Pareto
exponent suggests the rich really are rich and the poor are poor.
Notice that in the presently examined exchange market, one can
be rich either because of owning money or because of having many
apples. There is no king owning both.

We have also considered the case when   part of the money
disappears from the market under the form of ``taxes'',  at each
transaction. We have found that  many characteristics are quite
similar to the ''no tax'' case, but again the difference is mostly
seen in the distribution of wealth, i.e.,  the poor gets poorer and the
rich gets  richer. Most of the results appear ''quite natural''.

Notice also that recent studies of UK data show that
the Pareto exponent associated with the medium to high income range
has over the decade 1992 -2002 decreased from 3.3 to 2.7. During this
period taxation has increased\cite{Richmond3} .
It can be wondered \cite{Richmond3} if this represents the
impact of socialism since a higher slope value is associated with
greater redistribution of money, whereas a low value of the Pareto
exponent suggests the rich really are rich and the poor are poor.

   We are aware that in our simple model  we have much
reduced macro and micro economy conditions and therefore  further
modifications of our model may be needed. This however will reduce
the simplicity and transparency of the present model ({\it  Ockham
razor}).

In future work, it could be useful to discuss the ''differences'' between classical and
neoclassical theories  of {\it price/apple} and {\it value/money}. The
relevance  of  trade transactions  in which the
prices differ in time depending on the quantity of values/prices (or money/goods) to be exchanged
at a given time through (several) interacting agents, as in real economic societies
should be examined. The  relationship between trading  volume and  the prices   of  both financial
assets  and  commodities  is  also of  significant relevance  to
economists,   traders and politicians.
\vskip 0.6cm

{\bf Acknowledgments}

\vskip 0.6cm

MA thanks  C. Deissenberg, S. Hallegatte, R. Lambiotte and P.
Richmond for stimulating discussions and comments and acknowledges
support by the EC Project 'Extreme events: Causes  and
consequences (E2-C2), i.e.,   Contract No 12975 (NEST). AP thanks
KBN and CGRI for partial support.  This work was part of
investigations supported by the COST P10 ''Physics of  Risk'' STSM
program.

 \vskip 1cm

    %\vskip 2.6cm (*) email:   marcel.ausloos@ulg.ac.be \\

   % (#) email:   apekal@ift.uni.wroc.pl \\

\end{document}